\documentclass[twocolumn]{aastex631}
\usepackage{multirow}
\usepackage[tbtags]{amsmath}

\newcommand{\psr}{PSR~J2039$-$5617}
\newcommand{\fgl}{4FGL~J2039.5$-$5617}
\newcommand{\gr}{$\gamma$-ray}

\graphicspath{{./}{figures/}}

\begin{document}

\title{Revisiting $\gamma$-Ray Orbital Modulation in the Redback Millisecond Pulsar PSR~J2039$-$5617}

\author{Mengqing Zhang}
\affiliation{Department of Astronomy, School of Physics and Astronomy, Key Laboratory of Astroparticle Physics of Yunnan Province, Yunnan University, Kunming 650091, People's Republic of China; zhangpengfei@ynu.edu.cn}

\author{Shengbin Pei}
\affiliation{Department of Astronomy, School of Physics and Astronomy, Key Laboratory of Astroparticle Physics of Yunnan Province, Yunnan University, Kunming 650091, People's Republic of China; zhangpengfei@ynu.edu.cn}

\author{Shan Chang}
\affiliation{Department of Astronomy, School of Physics and Astronomy, Key Laboratory of Astroparticle Physics of Yunnan Province, Yunnan University, Kunming 650091, People's Republic of China; zhangpengfei@ynu.edu.cn}

\author{Pengfei Zhang}
\affiliation{Department of Astronomy, School of Physics and Astronomy, Key Laboratory of Astroparticle Physics of Yunnan Province, Yunnan University, Kunming 650091, People's Republic of China; zhangpengfei@ynu.edu.cn}

\begin{abstract}
\psr~is a redback millisecond pulsar binary system consisting of a compact star with
a mass of 1.1--1.6~M$_\sun$ and a low-mass companion of 0.15--0.22~M$_\sun$.
For this binary, we performed a timing analysis using 16 years of data from the Fermi Large Area Telescope, covering the period from 2008 August to 2024 October.
Our analysis detected an orbital modulation with a period of 0.2279781~days at
a significance level of $\sim4\sigma$, which is in good agreement with previous findings.
However, unlike previous reports, we identified a transition in the orbital modulation
around 2021 August, after which the orbital signal disappeared.
We speculate that the system may be undergoing a transition from a rotation-powered
to an accretion-powered state at this epoch.
Additionally, we conducted the phase-resolved and spectral analyses,
and in the phase-resolved results, we observed an anti-correlation between
its \gr~and X-ray emissions, which consistent with the predictions of high-energy
radiation models for such systems.
We provide some predictive discussions based on
the results of \gr~data analysis, and future Fermi-LAT observations will determine
whether these predictions hold true.
\end{abstract}

\section{Introduction}
\label{Intro}

Millisecond pulsars (MSPs) are generally regarded as the rapidly rotating neutron stars
that have undergone a recycling process. By accreting material from a companion star,
they gain angular momentum, which accelerates their rotation and reduces their
spin period to the millisecond range \citep{bv91}.
These pulsars are most commonly found in binary systems. Among them, two prominent
subclasses of eclipsing MSP binaries are the so-called spider systems:
Redbacks and Black Widows, both of which represent special categories of
compact binary evolution \citep{rob13,cct+13}.

Black Widow system typically consists of a neutron star and a very low-mass companion star,
usually $<$0.05 M$_\sun$, its orbital period is typically under 10 hours.
In Black Widow, the neutron star’s strong radiation and particle wind continuously strip mass
from the companion star, effectively shredding it over time. The most famous example is
PSR~B1957+20 \citep{fst88}, which was first identified as a Black Widow system.
As Black Widow, Redback systems contains a companion star with a mass ranging from 0.1 to
0.5~M$_\sun$. The orbital periods of these systems usually span from several hours to a few
days. The companion star in Redback systems is typically a low-mass main-sequence star or
a slightly evolved star. A well-known example of a Redback system is PSR J1023+0038,
which exhibits a remarkable behavior of transitioning between a millisecond pulsar
and a low-mass X-ray binary (LMXB) state, depending on the accretion activity \citep{asr+09}.
It is widely believed that Redback systems may be evolutionary precursors to
Black Widow systems \citep{bdh14}.
As the companion star in a Redback system continues to lose mass due to the pulsar's intense
radiation and wind, the system could eventually evolve into a Black Widow system,
with the companion becoming increasingly stripped and more negligible in mass \citep{acr+82}.

\psr~is a Redback MSP binary system with a spin period of 2.65~ms.
Its \gr~pulsations were first detected
in Fermi-LAT data through a directed search performed with the Einstein@Home
distributed volunteer computing system \citep{cnv+21}, and subsequent radio pulsations
were reported by \citet{cms+21}. Prior to these discoveries, \citet{smd+15}
identified its orbital period of 0.2245-day from X-ray and optical observations and
proposed that the system was the counterpart of the previously unassociated
\gr~source 3FGL~J2039.6$-$5618 detected by Fermi-LAT. Later, \citet[][hereafter as Ng2018]{nts+18}
analyzed Fermi-LAT observations spanning from 2008 August 4 to
2018 January 30. By applying the Rayleigh test, they found evidence for \gr~orbital
modulation between 2008 August 4 and 2015 January 18 (MJD 57,040);
however, this modulation appeared to vanish afterward.
They suggested that the intrabinary shock (IBS) in this system may be located closer to
the pulsar, leading to a lower Lorentz factor of the pulsar wind and consequently
weakening the inverse Compton (IC) radiation, rendering the orbital modulation
undetectable in \gr s.
Moreover, the transition of the redback system from the rotation-powered MSP state
to the accretion-powered LMXB state is crucial for understanding the evolution of
compact binary systems.
The disappearance of orbital modulation could signal such a transition,
marking a critical phase in the evolution of these systems.
If this epoch indeed represents the shift between these two states,
it would offer a unique opportunity to study the evolutionary pathway connecting them.
Motivated by this possibility, we have conducted a revised timing analysis
using Fermi-LAT data from \psr.

In \gr s, \psr~was first detected as an unidentified \gr~source in the 1FGL catalog,
designated 1FGL~J2039.4$-$5621 \citep{1fgl2010}. It was later listed as
2FGL~J2039.8$-$5620 and 3FGL~J2039.6$-$5618 in the 2FGL and 3FGL catalogs,
respectively \citep{2fgl2012,3fgl2015}. Subsequently,
\citet{smd+15} identified its X-ray and optical counterparts and classified
it as a candidate Redback MSP.
Based on 14 years of observations covering events in the
50~MeV--1~TeV, the source was reclassified as \fgl~\citep{4fgl-dr4}.
Here, we performed a timing analysis using $\sim$16~years of \gr~events of \fgl~from
Fermi-LAT, detecting an orbital period of 0.2279781~days with a significance
of $\sim4\sigma$. This period is consistent with the orbital parameters reported in the
pulsar ephemerides of \citet{cnv+21}, \citet{cms+21}, and \citet{ccc+22}.
Moreover, we found that its \gr~orbital modulation persisted steadily from 2008 August 4 to
2021 August 1, rather than ceasing in 2015 January 18 as reported by Ng2018.
The duration of the orbital period signal observed in the Fermi-LAT data has increased twofold.
The detailed data processing is described in Section~\ref{sec:lat-data},
followed by a summary and discussion in Section~\ref{sec:dis}.
The following numbers in parentheses represent the errors
associated with the last digit of the parameter values.

\section{Data Analysis and Results}
\label{sec:lat-data}

\subsection{Data Reduction And Best-fit Results}
\label{sec:model}
We selected Pass 8 Front+Back events with evclass~=~128 and evtype~=~3
within the energy range of 0.1--500.0 GeV, centered on \fgl~(R.A. = 309$^{\circ}$.8974 and Decl. =
$-$56$^{\circ}$.2836) and covering a $20^{\circ}\times20^{\circ}$ region of interest.
The observation period spans from 2008~August~4 to 2024~October~23 (MJD 54682.687--60606.955).
Events with zenith angles greater than 90$^{\circ}$ were excluded, and only high-quality
events from good time intervals were retained using the filter condition
“DATA\_QUAL~$>$~0 \&\& LAT\_CONFIG==1”. The diffuse \gr~emissions from the Galactic and
extragalactic isotropic components were modeled using the templates gll\_iem\_v07.fits
and iso\_P8R3\_SOURCE\_V2\_v1.txt, respectively. This data reduction and analysis
were conducted using Fermitools version~2.2.0.

%%%%%%%%%%%%
\begin{table}
\begin{center}
\caption{Best-fit results from likelihood analysis}
\begin{tabular}{ccccccc}
\hline\hline
Model & \multicolumn{5}{c}{Parameter valus} &  \\
\hline
 & $\gamma$ & $d$ & TS & $F_{\rm ph}$ & \\
\multirow{5}{*}{PLE4}   & 1.89(4) & 0.32(4) & 2818.74 & 1.61(10) & Whole\\
   & 1.87(4) & 0.35(4) & 2202.67 & 1.52(11) & P$_1$\\
   & 2.00(7) & 0.20(6) & 638.30  & 2.07(28) & P$_2$\\
   & 1.93(5) & 0.37(6) & 1693.29 & 1.92(16) & Ph$_{\rm ic}$ \\
   & 1.79(8) & 0.29(7) & 632.859  & 1.05(15) & Ph$_{\rm sc}$ \\
\hline
\end{tabular}
\label{tab:par}
\end{center}
{\bf Notes. }{Best-fit parameters of the likelihood, with $F_{\rm ph}$ in units of
              $\times10^{-8}$~photons~cm$^{-2}$~s$^{-1}$. The parameters $E_0$ and
              $b$ for the PLE4 model is fixed at $\sim$1.10~GeV and 2/3, respectively.}
\end{table}
%%%%%%%%%%%%
\begin{table*}
\begin{center}
\caption{Additional $\gamma$-ray point sources with their results}
\begin{tabular}{cccccc}
\hline\hline
New $\gamma$-ray sources & \multicolumn{1}{c}{Coordinate (error)$^a$} & $d^b$ & $\Gamma$ & TS & $F_{\gamma}^c$\\
\hline
& R.A. ~~~ Dec.~~~~~[radian] & [radian] & & & \\
\hline
NGS$_1$ & 308.147 $-51.870$ (0.065) & 4.533 & 2.12(20) & 38.83 & 6.30$\pm$1.10 \\
NGS$_2$ & 302.978 $-56.727$ (0.044) & 3.841 & 2.45(16) & 32.24 & 3.19$\pm$0.99 \\
NGS$_3$ & 310.749 $-57.936$ (0.003) & 1.714 & 1.64(14) & 41.89 & 0.32$\pm$0.17 \\
NGS$_4$ & 302.506 $-59.742$ (0.031) & 5.217 & 2.11(20) & 21.16 & 1.09$\pm$0.59 \\
NGS$_5$ & 304.726 $-53.316$ (0.034) & 4.205 & 2.21(15) & 27.87 & 1.86$\pm$0.70 \\
NGS$_6$ & 307.584 $-53.696$ (0.191) & 2.909 & 2.13(23) & 11.22 & 0.92$\pm$0.60 \\
\hline
\end{tabular}
\label{tab:ngsrc}
\end{center}
{\bf Notes. }{(a) Coordinates (J2000) and positional uncertainties of the six nearby NGSs,
              obtained using the tool \emph{gtfindsrc}. (b) Angular distances of the six NGSs from
              the target source. (c) Integrated photon fluxes in 0.1--500.0~GeV with units
              of $\times$10$^{-9}$~photons~cm$^{-2}$~s$^{-1}$.}
\end{table*}
%%%%%%%%%%%%
\begin{figure*}
\centering
\includegraphics[angle=0,scale=0.66]{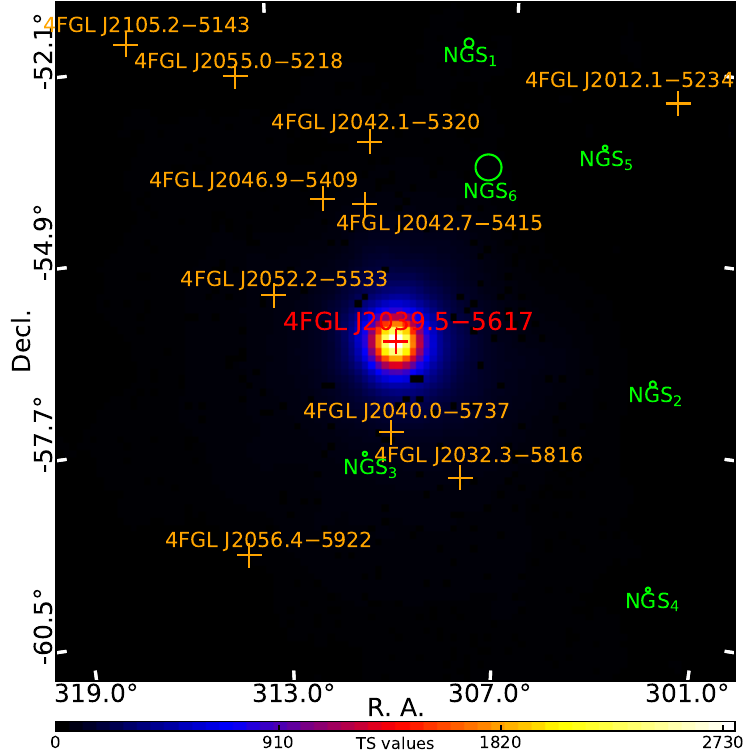}
\includegraphics[angle=0,scale=0.66]{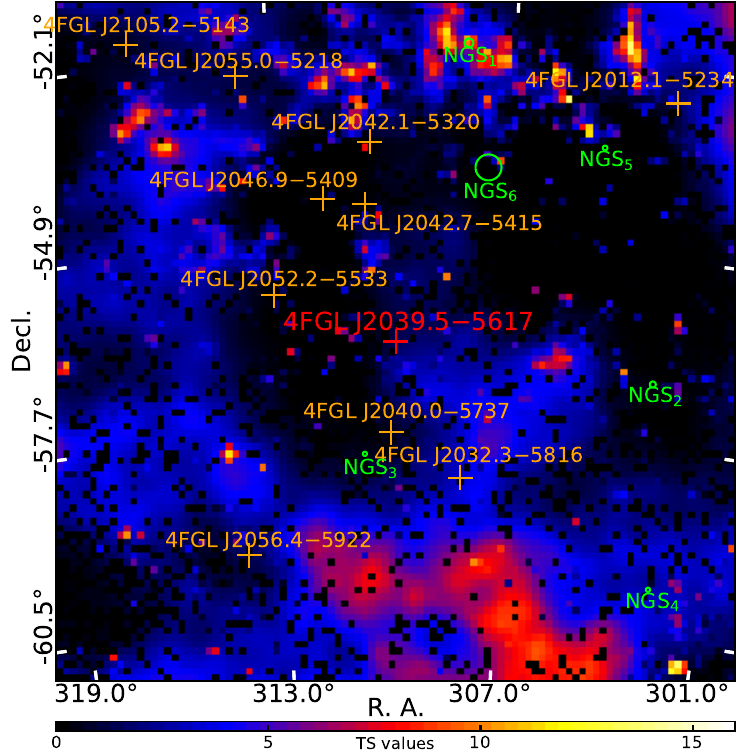}
\caption{0.1--500.0~GeV TS maps centered on \fgl~with a region of $10^{\circ}\times10^{\circ}$.
         The target is marked by a red crosse, while other \gr~sources from the 4FGL
         catalog are denoted by orange crosses, the six NGSs are shown as
         green circles, with radii proportional to their positional uncertainties.
         Left panel: TS map showing the \gr~emissions
         from \fgl, derived from the best-fit model with the target excluded.
         Right panel: Residual TS map generated using the same model, but including \fgl.}
\label{fig:tsmap}
\end{figure*}
%%%%%%%%%%%%

\begin{figure*}
\centering
\includegraphics[angle=0,scale=0.58]{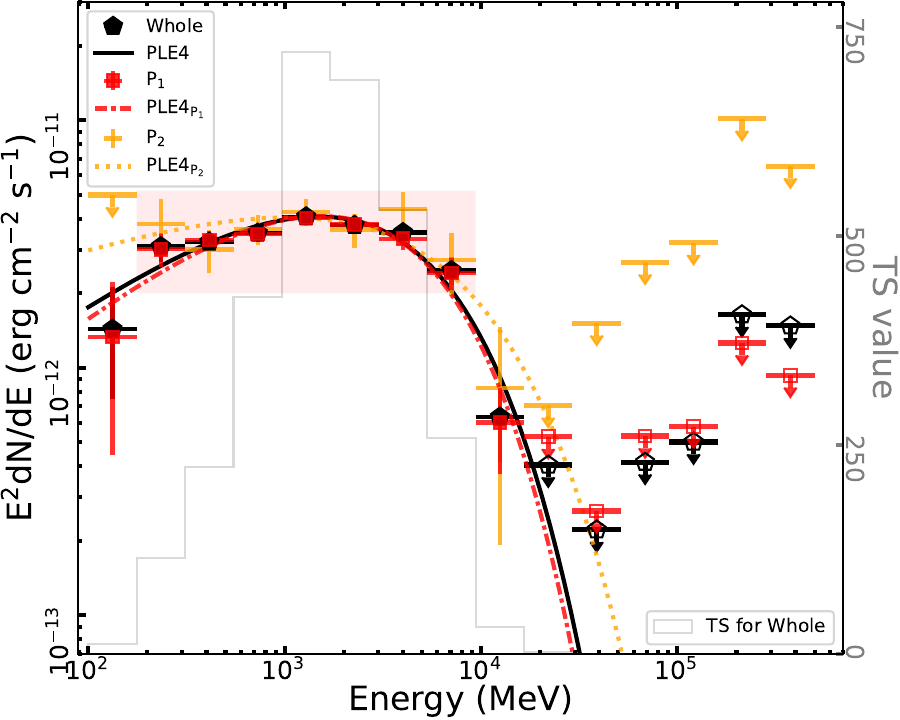}
\includegraphics[angle=0,scale=0.58]{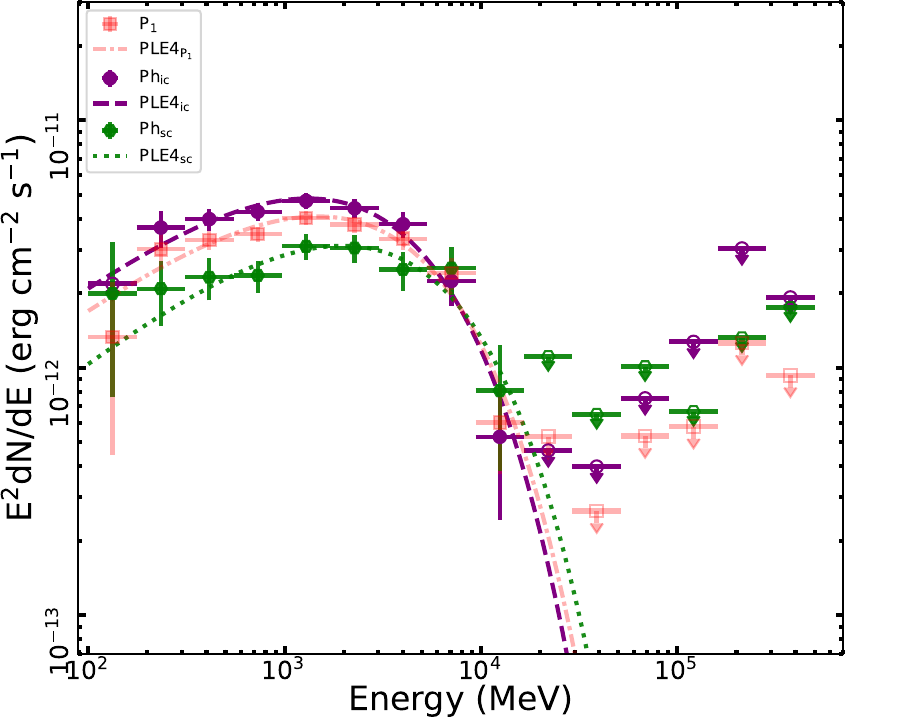}
\caption{SEDs of \fgl~in 0.1--500.0~GeV. Left: SEDs for three time intervals, 2008-08-04
         to 2024-10-23 (black, Whole), 2008-08-04 to 2021-08-01 (red, P$_1$),
         and 2021-08-01 to 2024-10-23 (orange, P$_2$), with their best-fit PLE4 models
         shown as a black solid, red dashed-dotted, and orange dotted line, respectively.
         For the Whole results, TS values for each energy bin are displayed as a gray histogram
         (on right y-axis).
         Right: SEDs during the epoch where the AP light curve shows periodicity (P1).
         The red points match those in the left panel, while purple and green points
         correspond to phase intervals around inferior and superior conjunction,
         respectively, as that indicated by the pink and cyan shaded regions in
         Figure~\ref{fig:phlc}. Their best-fit models are shown as purple dashed and
         green dotted lines, respectively.}
\label{fig:sed}
\end{figure*}

In the 4FGL, \fgl's \gr~spectral shape is represented with a new subexponentially
cutoff power-law (PLE4). Using 16 yr data, we updated its parameters by performing a binned
maximum likelihood analysis based on a model file from 4FGL.
The best-fit parameters were saved in an updated model file. To identify significant potential
\gr~sources not included in 4FGL, we generated a test statistic (TS) map of \gr~residuals
after excluding emissions from all 4FGL sources in the model. The residual map revealed
several new, relatively significant \gr~excesses.
To account for these excesses, we iteratively added new \gr~point sources (NGSs),
with spectral shape of PowerLaw, to the
model file and ultimately identified six additional sources. Their \gr~positions
were determined using \emph{gtfindsrc}. We then re-performed the data analysis to update
the model and saved the results as a best-fit model file.
Based on this model, we generated two TS maps for \fgl,
as shown in Figure~\ref{fig:tsmap}.
The left panel displays the \gr~emissions from \fgl, and the right panel shows
the residual \gr s after excluding emissions from all sources in the best-fit model.
In the residual map, the maximum TS value among all pixels is $\sim$13.5.
The best-fit results for the target, labeled “Whole”, are summarized in Table~\ref{tab:par}
and shown in Figure~\ref{fig:sed} with a black solid line.
While the results for the six NGSs are presented in Table~\ref{tab:ngsrc}.
These results indicate that the \gr~emissions from \fgl~are better described by
the PLE4 model, and subsequent data analysis was conducted based on the best-fit model.

%%%%%%%%%%%%
%%%%%%%%%%%%

\subsection{Spectral Analysis}
For the spectral analysis, we used Fermi-LAT events in the 0.1--500.0~GeV energy range.
The spectral energy distribution (SED) of \fgl~was extracted by dividing this range into
15 equally logarithmically spaced energy bins, and a binned likelihood analysis was
performed in each bin to derive the \gr~flux. In this procedure, only the normalization
parameters of \gr~sources within 5$^\circ$ of the target were left free,
while all other parameters were fixed to the values obtained from the new model file.
For energy bins with TS values $\ge$ 4, the fluxes are shown as black data points in
Figure~\ref{fig:sed}, whereas for bins with TS $<$ 4, 95\% flux upper limits are plotted.
TS values for each energy bin are shown with a gray histogram on the right y-axis.
As shown in Figure~\ref{fig:sed}, the target is clearly significant in the 0.18--9.39~GeV range,
where all bins have TS values $>$ 100. Accordingly, the subsequent timing analysis
was primarily concentrated on this energy range.

%%%%%%%%%%%%
\begin{figure*}
\centering
\includegraphics[angle=0,scale=0.7]{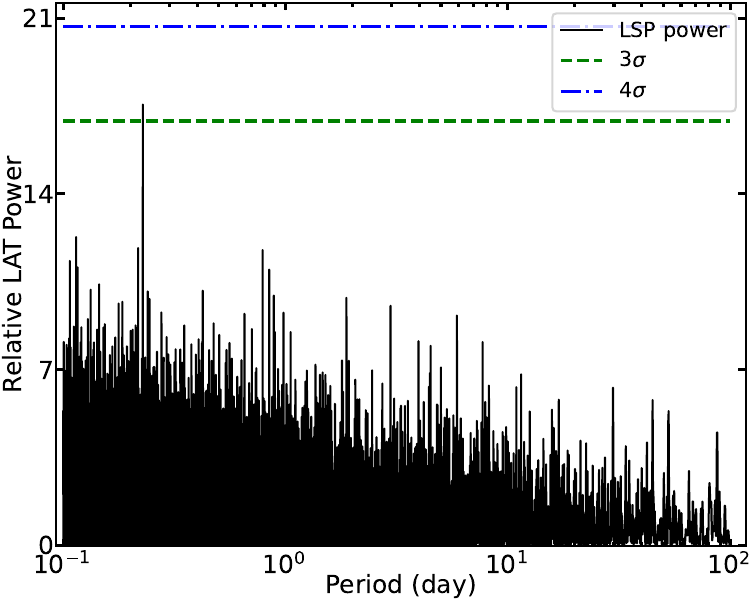}
\includegraphics[angle=0,scale=0.7]{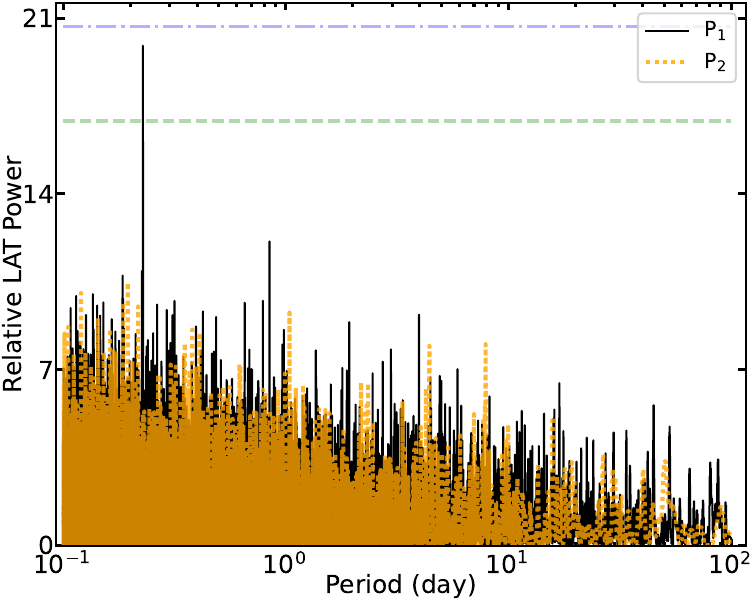}
\caption{LSP power spectra (black histogram) of \fgl~derived from the AP light curve in 0.18--9.39~GeV.
         Left: LSP power spectrum for the Whole AP light curve, with the 3$\sigma$
         and 4$\sigma$ confidence levels indicated
         by green dashed and blue dash-dotted lines, respectively.
         Right: LSP power spectra for the AP light curve in P$_1$ and P$_2$,
         shown in black and orange, respectively.}
\label{fig:lsps}
\end{figure*}
%%%%%%%%%%%%
\begin{figure}
\centering
\includegraphics[angle=0,scale=0.68]{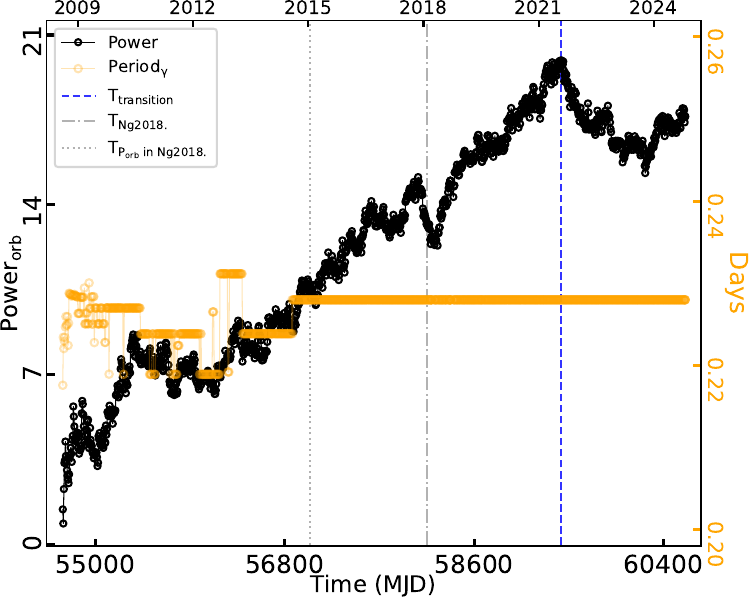}
\caption{LSP power values around the periodic signal obtained from the accumulated data.
         The time axis corresponds to the end dates of the AP light curves, all starting
         from the same epoch (MJD~54682.687). As an example, the AP light curve spanning
         from 2008 August 4 to 2021 August 1 yields the highest LSP power of $\sim$19.95
         (blue dashed line). The gray dashed-dotted line indicates the end date of
         the dataset analyzed by Ng2018, while the gray dotted line marks
         the epoch when orbital modulation was reported to disappear in their work.
         In comparison, the blue dashed line denotes the epoch at which we find
         the modulation to vanish. The evolution of the signal period with
         accumulated data is plotted as orange circles on the right y-axis.}
\label{fig:pvst}
\end{figure}

%\subsection{Light Curve Construction and Timing Analysis}
\subsection{Timing Analysis}
\label{sec:timing}

To conduct a blind search for the orbital period of \psr~in the GeV, we applied a modified
aperture photometry (AP) method to construct the light curve with a time bin of 500~s.
%176.44038643562388, 9392.560525107727 MeV
To optimize the signal-to-noise ratio for detecting periodic signals,
we focused on the energy range of 0.18--9.39~GeV, corresponding to the pink-shaded
region in the SED, where the target is most significant. Testing different energy cuts could
slightly increase the power of the detected peak,
but we did not investigate how the power of the periodic signal depends on the energy range.
For our timing analysis, events centered on
\fgl~within an aperture radius of 2.76$^\circ$ were selected, following the criteria
established by \citet{abd+10}. Using \emph{gtmktime}, we excluded periods when the target was
within 5$^\circ$ of the Sun or Moon to eliminate their potential influence. To minimize the
impact of significant exposure variations across time bins, exposures were calculated using
\emph{gtexposure}. Probabilities for each event, indicating their likelihood of originating
from \fgl, were assigned using \emph{gtsrcprob} based on the new model file. We then
constructed the AP light curve, using the assigned probabilities as
weights \citep{k11,lat+12,cor+19}.
Finally, the times of light curve were barycenter corrected using \emph{gtbary}.

We used a common timing analysis tool, the Lomb-Scargle Periodogram (LSP), to obtain the
power spectrum of the AP light curve \citep{l76,s82,zk09}. Compared to other algorithms,
the LSP is highly effective in detecting periodic signals in unevenly sampled data. It
handles noisy data well, is computationally efficient, and provides clear and reliable results
for identifying and analyzing periodicities. We searched for periodic signals within a
frequency range of $f_{\rm min}$~=~0.01 to $f_{\rm max}$~=~10 day$^{-1}$ with a frequency
resolution of $\delta f=1/T{_{\rm obs}}$, where $T{_{\rm obs}}\sim5924.268$~day is the
duration of Fermi-LAT observations.
The number of independent frequencies (the trial factor) was calculated as
$N=(f_{\rm max} - f_{\rm min})/\delta f$ = 59,183.
To be conservative, we adopted this trial numbers for all subsequent timing analyses.
The LSP results are shown in the left
panel of Figure~\ref{fig:lsps}. A distinct power peak, with a power value of $\sim$17.64,
was observed around the orbital period reported in previous
literatures \citep{smd+15,nts+18,cnv+21,cms+21}. We determined that the probability
($p_{_{\rm lsp}}$) of obtaining this power level due to random fluctuations (Gaussian white
noise) is approximately $2.17\times10^{-8}$ \citep{l76,s82}. After accounting for the number
of trials $N$, we calculated a false alarm probability ($FAP$) to be
$FAP=1-(1-p_{_{\rm lsp}})^N \sim N\times p_{_{\rm lsp}} = 1.28\times10^{-3}$,
corresponding to a confidence level of 3.2$\sigma$. In Figure~\ref{fig:lsps},
we also show 3$\sigma$ and 4$\sigma$ confidence levels with green dashed and
blue dashed–dotted lines, respectively.

Considering the orbital modulation transitions in \gr s for the Redback system, as reported
by Ng2018 and \citet{ccc+22}, we analyzed the time-dependent variations in the
power values of this periodic signal. The power values are shown as black
circles in Figure~\ref{fig:pvst}. Additionally, the temporal changes in the period of the
orbital modulation are colored in orange on the right side of Figure~\ref{fig:pvst}.
As shown in the figure, the orbital modulation transition occurs at 2021 August 1
(MJD~59427.683, defined as $T_{\rm transition}$), marked by a blue dashed line.
Based on this transition, we divided the Fermi-LAT observations into two segments:
data prior to $T_{\rm transition}$ is designated as Part~1 (P$_1$), and data after
$T_{\rm transition}$ as Part~2 (P$_2$).

Therefore, we divided the AP light curve into two parts with using the division time of $T_{\rm transition}$,
and generated their power spectra using the same timing analysis process, they are displayed
in the right panel of Figure~\ref{fig:lsps}.
For P$_1$ (the black solid histogram), the orbital signal power rises to
$\sim$19.95, with a FAP of $\sim1.27\times10^{-4}$, corresponding to a 3.8$\sigma$
confidence level, based on the previous estimation method.
And its period is $P_{\rm orb}=0.2279781(9)$~days, where the uncertainties were calculated by
$\delta P=\frac{3}{8}\frac{P^2}{T_{\rm obs}\sqrt{p_{_{\rm n}}}}$ \citep{ccc+22}.
In contrast, for P$_2$ (orange dashed histogram),
the orbital modulation signal has completely disappeared.
Our results show some discrepancies compared to those reported by Ng2018.

%%%%%%%%%%%%
\begin{figure}
\centering
\includegraphics[angle=0,scale=0.65]{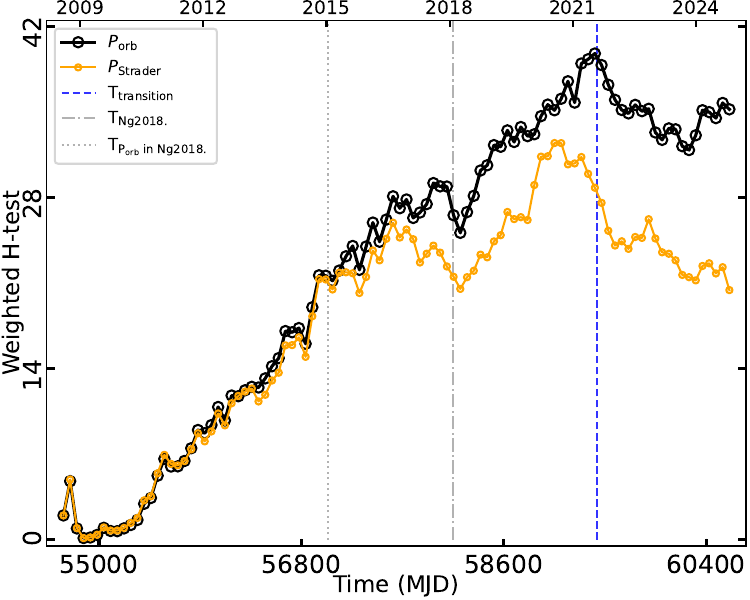}
\caption{Cumulative H-test results obtained using the orbital periods $P_{\rm orb}$
         (derived in this work, black) and $P_{\rm Strader}$ (adopted by Ng2018,
         orange). All others are the same as in Figure~\ref{fig:pvst}.}
\label{fig:cca}
\end{figure}
%%%%%%%%%%%%

To further investigate the reasons for the discrepancy,
we employed the H-test method, which is a statistical
test similar to the Rayleigh test, to our data. The results are shown in Figure~\ref{fig:cca}.
We performed the H-test using both the orbital period of $P_{\rm orb}$ and $P_{\rm Strader}$ (=0.2279817~days)
that adopted by Ng2018. As shown in Figure~\ref{fig:cca}, the difference between
this two results begins to increase gradually around MJD~57040 (gray dotted line).

We speculate that this difference mainly arises from the use of different orbital periods.
When the adopted orbital period deviates from the true value, the impact of
this deviation on the results is relatively small in the early stage of observation
(e.g., before MJD~57040). However, as time progresses, the cumulative error in
the orbital period increases, resulting in a more significant effect on the results
in the later stage (after MJD~57040). In addition, compared with Ng2018,
we used updated Fermi-LAT data and an improved version of Fermi-Tools. And the 4FGL-DR3 also
includes several newly identified \gr~sources (e.g., 4FGL~J2046.9+5409), and the nearby source,
3FGL~J2051.8$–$5535, also exhibited a \gr~flare around MJD 57040 (see the Figure 9 in Ng2018).
These factors may have increased the probability of misidentifying non-target \gr s as originating
from the target. Such non-target \gr s, which are not modulated, could weaken the significance
of the periodic signal in the data,
thereby contributing to the differences in this timing results between ours and Ng2018.

Based on $P_{\rm orb}$, the H-test results show that the periodic signal reaches its maximum at
MJD~59409.865 (around $T_{\rm transition}$),
which is consistent with the LSP results. The H-test value for the whole data is 35.22,
corresponding to a $p$-value of $\sim7.61\times10^{-7}$ ($5.0\sigma$).
And the maximum H-test value is 39.79, has a $p$-value of $\sim1.22\times10^{-7}$ ($5.3\sigma$).
When calculated H-test results using the period of $P_{\rm Strader}$, the signal becomes
similar to that reported by Ng2018 over the same time interval.

We then performed likelihood analysis separately on each of these two data
segments, and the results are summarized in Table~\ref{tab:par}, labeled as P$_1$ and P$_2$,
respectively. We also created their SEDs and show them in left panel of Figure~\ref{fig:sed}.
From these results, we observe that the spectral shape of
the target shows no significant differences between the two segments.

%%%%%%%%%%%%
\begin{figure}
\centering
\includegraphics[angle=0,scale=0.96]{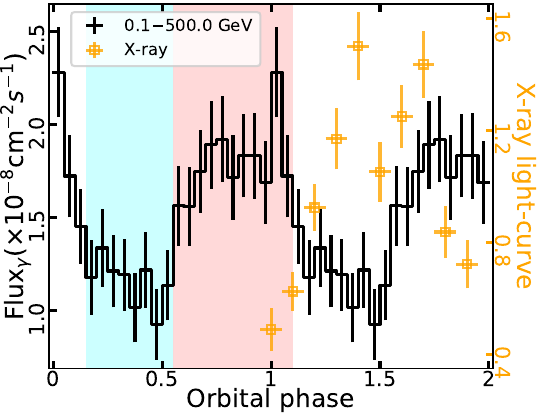}
\caption{Phase-resolved light curves derived from the 0.1--500.0~GeV Fermi-LAT events
         in P$_1$, folded over the \gr~orbital period of 0.2279781~days. The phase intervals
         around the inferior and superior conjunctions are indicated by the pink and
         cyan shades. The orange data points (right y-axis) show the schematic X-ray
         modulation profile of the counterpart to \psr, adapted from \citet{smd+15}.}
\label{fig:phlc}
\end{figure}
%%%%%%%%%%%%

\subsection{Phase-resolved Analysis}
\label{sec:phlc}
To investigate the \gr~orbital modulation profile of \psr, we performed a
likelihood analysis on the phase-resolved Fermi-LAT data, adopting
its orbital period of $P_{\rm orb}$ in the 0.1--500~GeV energy range.
The resulting phase-resolved likelihood light curve is presented in
Figure~\ref{fig:phlc}. These results clearly show that the \gr~flux of
\psr~is modulated by its orbital period. The modulation profile exhibits
a rapid rise followed by a gradual decline, and a possible structure with
two asymmetric peaks was detected in the phase-resolved light curve.

To explore potential variations in the \gr~spectral properties across different
orbital phases, we derived the SEDs around the inferior and superior conjunctions,
corresponding to phase ranges of 0.55--1.1 (denoted as Ph$_{\rm ic}$) and 0.15--0.55
(Ph$_{\rm sc}$), respectively.
A likelihood analysis was performed on events within these two phase ranges,
with the best-fit parameters summarized in Table~\ref{tab:par}. Based on these fits,
the SEDs were constructed and are shown in the right panel of Figure~\ref{fig:sed},
in purple and green, respectively. The results indicate that, aside from
a significant flux difference at lower energies (in $\sim$0.1--5.3~GeV),
the spectral shapes show no noticeable differences between the two orbital states.

To facilitate the comparison of the modulation profiles between \gr s and X-rays,
we extracted the schematic X-ray modulation profile of the counterpart to
\psr~from Figure~3 of \citet{smd+15} and replotted it in Figure~\ref{fig:phlc}
using the orange data points on the right y-axis. It can be seen that
the modulation profiles of \gr s and X-rays appear to be offset by $\sim$0.5 in phase.

\section{Summary and Discussion}
\label{sec:dis}
\psr~is a redback MSP binary system, consisting of a compact star with a mass
of 1.1--1.6~M$_\sun$ and a low-mass companion of 0.15--0.22~M$_\sun$ \citep{cnv+21,cms+21}.
\citet{smd+15} first identified its orbital period of 0.2245~days from both X-ray
and optical observations. Then Ng2018 reported an evidence of
\gr~orbital modulation from 2008 August 4 to 2015 January 18,
they also provided a detailed discussion for the \gr~emissions from this binary system.
In this work, we extend the analysis to Fermi-LAT data spanning from 2008 August 4 to 2024 October 23. Our results confirm the orbital period
at 0.2279781~days and suggest that the modulation persists longer than previously reported,
disappearing around 2021 August.
The observational time span over which the orbital
period of \psr~was detected is approximately twice as long as
that reported by Ng2018.

A comparison between our results and those of Ng2018 highlights an interesting
discrepancy in the reported significance of the orbital period signal.
While Ng2018 found a $\sim4\sigma$ confidence level, our analysis, despite being
based on nearly twice as much data, yields a similar significance.
One possible explanation lies in the different statistical methods adopted,
Ng2018 employed the Rayleigh test, whereas we applied the LSP.
Alternatively, the discrepancy may arise from differences in how trial factors
were handled in our analysis, as including a larger number of trials naturally
reduces the estimated significance.
When applying an H-test approach similar to that adopted by Ng2018,
the signal significance increased to 5.3$\sigma$.
This suggests that the robustness of the periodic signal detection is highly
sensitive to both the choice of statistical method and the treatment of trial
corrections.

A comparison between our Figure~\ref{fig:phlc} and the \gr~folded light curves
presented in Ng2018 (i.e., their Figures~4 and~5) reveals clear differences
between the results obtained from photon folding and those derived from
likelihood analysis when constructing the phase-resolved light curve.
Similar conclusions are observed in other cases we examined (Zhang et al., in preparation).
From experience with data analysis, the likelihood-based approach is generally
more reliable for certain binaries, particularly for relatively faint \gr~sources.
While for very bright \gr~binaries, such as LS 5039 \citep{ls5039+09},
LS I +61 303 \citep{lsi61303+09}, and 1FGL~J1018.6$-$5856 \citep{cor+16}
the two methods are expected to yield results that are largely consistent.
Another possible explanation lies in the difference in data volume mentioned
above: Ng2018 used only about six years of data to construct the folded light curve,
whereas our phase-resolved analysis is based on $\sim$13 years of data,
nearly doubling the dataset compared with theirs.

The phase-resolved results around the superior (indicated by the cyan shade
in Figure~\ref{fig:phlc}) and inferior (shown in pink) conjunctions indicate that
the \gr~spectra of the source remain nearly unchanged, with only the flux varying in 0.1--5.3~GeV,
likely due to geometric factors. If the \gr~emission mechanism is based on the
IC scattering between the pulsar wind and the stellar soft photons,
this suggests that the soft photon density around the stellar and
the relativistic charged particles from the pulsar wind in the emitting region
remain stable. Therefore, the spectral properties of the \gr~emissions do not
modulate with the orbital phase. The phase-resolved light curve further shows
that the target emits \gr s throughout the entire orbital phase, suggesting
that the \gr~emission region in the IC model has a significantly larger extent,
meaning that the emission is spread out over a wide area rather than being
concentrated in a narrow beam.
This larger emission region allows an observer
to detect the emission in any geometric configuration. From this perspective,
it also resembles the \gr~magnetosphere radiation mechanism in pulsars \citep{rmh10}.

The observed phase relationship between the X-ray and \gr~modulations provides
important insight into the emission geometry of this system.
While \gr~maxima are typically associated with inferior conjunction of orbital phase
(when the companion star passes in front of the pulsar from the observer’s perspective)
and X-ray maxima with superior conjunction, the limited observational coverage prevents
us from establishing the precise orbital phases for the target.
Nevertheless, the $\sim$0.5 phase offset between the \gr~and X-ray light curves
(shown in Figure~\ref{fig:phlc})
is consistent with the expected anti-correlation predicted by theoretical models
of IC scattering and relativistic Doppler boosting in compact binaries.
This agreement supports the interpretation that the observed high-energy radiation
arises from the interplay between these mechanisms.
This anti-correlated behavior between \gr s and X-rays has been reported in
numerous Redback and Black Widow pulsars, including the recent case reported
by \citet{saw24}, as well as in \gr~binaries \citep{cor+16,cor+19}.
Our results therefore not only align with these earlier studies but also extend
the observational evidence to this newly examined source.
Conducting detailed phase-resolved analyses will be critical to tightly constrain
the orbital geometry and to assess whether the source conforms to
the canonical phenomenology of Redback and Black Widow pulsars.

Since we expect the \gr~orbital modulation to originate from
the geometric configuration of IC scattering, the modulation profile
is typically anticipated to exhibit a symmetric single-peak structure \citep{Wu2012}.
However, the phase-resolved light curve of \psr~reveals a possible structure with
two asymmetric peaks. Currently, the statistical significance is still
insufficient for an in-depth discussion. Future Fermi-LAT observations,
with larger datasets, may provide the necessary data to either confirm
or refute the presence of this structure, thereby offering valuable
insights into the emission mechanisms operating in spider pulsar systems.

Last but not least, we know that the disappearance of periodic modulation may mark
a critical phase in the evolution of compact binaries, possibly corresponding to
the transition from a rotation-powered MSP state to an accretion-powered LMXB state.
We observed that, prior to 2021 August, the \gr~orbital modulation persisted,
suggesting that \psr~was in the MSP state during this period. Subsequently,
the system transitioned to the LMXB state, where the accretion process became dominant,
forming an accretion disk, disrupting the IBS, and preventing particles from
acquiring sufficient energy. This disruption in the IC scattering between the pulsar
and its companion star led to the interruption of orbital modulation in \gr s.
However, current observational evidence is insufficient to confirm this scenario.
To validate it, additional observational data are required, particularly optical
and/or X-ray data, to compare with previous observations and investigate
the presence of an accretion disk as well as variations in the X-ray luminosity
of \psr. Nevertheless, although \psr~may not have transitioned to an LMXB state,
some changes in the system did occur.

\begin{acknowledgments}
We would like to thank the anonymous referee for the helpful suggestions.
This work was partially supported by the National Natural Science Foundation
of China under grant Nos.~12233006, 12163006, and 12103046,
and by the Scientific Research and Innovation Project of Postgraduate Students
in the Academic Degree of Yunnan University Nos.~KC-242410143 and KC-252512090.
P.Z. and S.C. acknowledge support from the Xingdian Talent Support Plan - Youth Project.
\end{acknowledgments}

\bibliographystyle{aasjournal}
\bibliography{aas}

\begin{thebibliography}{}
\expandafter\ifx\csname natexlab\endcsname\relax\def\natexlab#1{#1}\fi
\providecommand{\url}[1]{\href{#1}{#1}}
\providecommand{\dodoi}[1]{doi:~\href{http://doi.org/#1}{\nolinkurl{#1}}}
\providecommand{\doeprint}[1]{\href{http://ascl.net/#1}{\nolinkurl{http://ascl.net/#1}}}
\providecommand{\doarXiv}[1]{\href{https://arxiv.org/abs/#1}{\nolinkurl{https://arxiv.org/abs/#1}}}

\bibitem[{{Abdo} {et~al.}(2009{\natexlab{a}}){Abdo}, {Ackermann}, {Ajello},
  {Atwood}, {Axelsson}, {Baldini}, {Ballet}, {Barbiellini}, {Bastieri},
  {Baughman}, {Bechtol}, {Bellazzini}, {Berenji}, {Blandford}, {Bloom},
  {Bonamente}, {Borgland}, {Bregeon}, {Brez}, {Brigida}, {Bruel}, {Burnett},
  {Buson}, {Caliandro}, {Cameron}, {Caraveo}, {Casandjian}, {Cavazzuti},
  {Cecchi}, {{\c{C}}elik}, {Chaty}, {Chekhtman}, {Cheung}, {Chiang}, {Ciprini},
  {Claus}, {Cohen-Tanugi}, {Cominsky}, {Conrad}, {Corbel}, {Corbet}, {Cutini},
  {Dermer}, {de Angelis}, {de Palma}, {Digel}, {Silva}, {Drell}, {Dubois},
  {Dubus}, {Dumora}, {Farnier}, {Favuzzi}, {Fegan}, {Focke}, {Fortin},
  {Frailis}, {Fukazawa}, {Funk}, {Fusco}, {Gargano}, {Gasparrini}, {Gehrels},
  {Germani}, {Giebels}, {Giglietto}, {Giordano}, {Glanzman}, {Godfrey},
  {Grenier}, {Grondin}, {Grove}, {Guillemot}, {Guiriec}, {Hanabata}, {Harding},
  {Hayashida}, {Hays}, {Hill}, {Horan}, {Hughes}, {Jackson}, {J{\'o}hannesson},
  {Johnson}, {Johnson}, {Johnson}, {Kamae}, {Katagiri}, {Kataoka}, {Kawai},
  {Kerr}, {Kn{\"o}dlseder}, {Kocian}, {Kuehn}, {Kuss}, {Lande}, {Larsson},
  {Latronico}, {Lemoine-Goumard}, {Longo}, {Loparco}, {Lott}, {Lovellette},
  {Lubrano}, {Madejski}, {Makeev}, {Marelli}, {Mazziotta}, {McEnery}, {Meurer},
  {Michelson}, {Mitthumsiri}, {Mizuno}, {Moiseev}, {Monte}, {Monzani},
  {Morselli}, {Moskalenko}, {Murgia}, {Nolan}, {Norris}, {Nuss}, {Ohsugi},
  {Omodei}, {Orlando}, {Ormes}, {Ozaki}, {Paneque}, {Panetta}, {Parent},
  {Pelassa}, {Pepe}, {Pesce-Rollins}, {Piron}, {Porter}, {Rain{\`o}}, {Rando},
  {Ray}, {Razzano}, {Rea}, {Reimer}, {Reimer}, {Reposeur}, {Ritz}, {Rochester},
  {Rodriguez}, {Romani}, {Roth}, {Ryde}, {Sadrozinski}, {Sanchez}, {Sander},
  {Saz Parkinson}, {Scargle}, {Sgr{\`o}}, {Sierpowska-Bartosik}, {Siskind},
  {Smith}, {Smith}, {Spandre}, {Spinelli}, {Strickman}, {Suson}, {Tajima},
  {Takahashi}, {Takahashi}, {Tanaka}, {Tanaka}, {Thayer}, {Thompson},
  {Tibaldo}, {Torres}, {Tosti}, {Tramacere}, {Uchiyama}, {Usher}, {Vasileiou},
  {Venter}, {Vilchez}, {Vitale}, {Waite}, {Wallace}, {Wang}, {Winer}, {Wood},
  {Ylinen}, \& {Ziegler}}]{ls5039+09}
{Abdo}, A.~A., {Ackermann}, M., {Ajello}, M., {et~al.} 2009{\natexlab{a}},
  \apjl, 706, L56, \dodoi{10.1088/0004-637X/706/1/L56}

\bibitem[{{Abdo} {et~al.}(2009{\natexlab{b}}){Abdo}, {Ackermann}, {Ajello},
  {Atwood}, {Axelsson}, {Baldini}, {Ballet}, {Barbiellini}, {Bastieri},
  {Baughman}, {Bechtol}, {Bellazzini}, {Berenji}, {Blandford}, {Bloom},
  {Bonamente}, {Borgland}, {Bregeon}, {Brez}, {Brigida}, {Bruel}, {Burnett},
  {Caliandro}, {Cameron}, {Caraveo}, {Casandjian}, {Cavazzuti}, {Cecchi},
  {{\c{C}}elik}, {Charles}, {Chaty}, {Chekhtman}, {Cheung}, {Chiang},
  {Ciprini}, {Claus}, {Cohen-Tanugi}, {Cominsky}, {Conrad}, {Corbel}, {Corbet},
  {Cutini}, {Dermer}, {de Angelis}, {de Luca}, {de Palma}, {Digel}, {Dormody},
  {do Couto e Silva}, {Drell}, {Dubois}, {Dubus}, {Dumora}, {Farnier},
  {Favuzzi}, {Fegan}, {Focke}, {Frailis}, {Fukazawa}, {Funk}, {Fusco},
  {Gargano}, {Gasparrini}, {Gehrels}, {Germani}, {Giebels}, {Giglietto},
  {Giordano}, {Glanzman}, {Godfrey}, {Grenier}, {Grondin}, {Grove},
  {Guillemot}, {Guiriec}, {Hanabata}, {Harding}, {Hayashida}, {Hays}, {Hill},
  {Hughes}, {J{\'o}hannesson}, {Johnson}, {Johnson}, {Johnson}, {Johnson},
  {Kamae}, {Katagiri}, {Kataoka}, {Kawai}, {Kerr}, {Kn{\"o}dlseder}, {Kocian},
  {Kuehn}, {Kuss}, {Lande}, {Larsson}, {Latronico}, {Longo}, {Loparco}, {Lott},
  {Lovellette}, {Lubrano}, {Madejski}, {Makeev}, {Marelli}, {Mazziotta},
  {McEnery}, {Meurer}, {Michelson}, {Mitthumsiri}, {Mizuno}, {Monte},
  {Monzani}, {Morselli}, {Moskalenko}, {Murgia}, {Nolan}, {Nuss}, {Ohsugi},
  {Okumura}, {Omodei}, {Orlando}, {Ormes}, {Paneque}, {Panetta}, {Parent},
  {Pelassa}, {Pepe}, {Pesce-Rollins}, {Piron}, {Porter}, {Rain{\`o}}, {Rando},
  {Ray}, {Razzano}, {Rea}, {Reimer}, {Reimer}, {Reposeur}, {Ritz}, {Rochester},
  {Rodriguez}, {Romani}, {Ryde}, {Sadrozinski}, {Sanchez}, {Sander}, {Saz
  Parkinson}, {Scargle}, {Sgr{\`o}}, {Shaw}, {Sierpowska-Bartosik}, {Siskind},
  {Smith}, {Smith}, {Spandre}, {Spinelli}, {Striani}, {Strickman}, {Suson},
  {Tajima}, {Takahashi}, {Takahashi}, {Tanaka}, {Thayer}, {Thayer}, {Thompson},
  {Tibaldo}, {Torres}, {Tosti}, {Tramacere}, {Uchiyama}, {Usher}, {Vasileiou},
  {Vilchez}, {Vitale}, {Waite}, {Wang}, {Winer}, {Wood}, {Ylinen}, \&
  {Ziegler}}]{lsi61303+09}
---. 2009{\natexlab{b}}, \apjl, 701, L123, \dodoi{10.1088/0004-637X/701/2/L123}

\bibitem[{{Abdo} {et~al.}(2010{\natexlab{a}}){Abdo}, {Ackermann}, {Ajello},
  {Allafort}, {Antolini}, {Atwood}, {Axelsson}, {Baldini}, {Ballet},
  {Barbiellini}, {Bastieri}, {Baughman}, {Bechtol}, {Bellazzini}, {Belli},
  {Berenji}, {Bisello}, {Blandford}, {Bloom}, {Bonamente}, {Bonnell},
  {Borgland}, {Bouvier}, {Bregeon}, {Brez}, {Brigida}, {Bruel}, {Burnett},
  {Busetto}, {Buson}, {Caliandro}, {Cameron}, {Campana}, {Canadas}, {Caraveo},
  {Carrigan}, {Casandjian}, {Cavazzuti}, {Ceccanti}, {Cecchi}, {{\c{C}}elik},
  {Charles}, {Chekhtman}, {Cheung}, {Chiang}, {Cillis}, {Ciprini}, {Claus},
  {Cohen-Tanugi}, {Conrad}, {Corbet}, {Davis}, {DeKlotz}, {den Hartog},
  {Dermer}, {de Angelis}, {de Luca}, {de Palma}, {Digel}, {Dormody}, {Silva},
  {Drell}, {Dubois}, {Dumora}, {Fabiani}, {Farnier}, {Favuzzi}, {Fegan},
  {Ferrara}, {Focke}, {Fortin}, {Frailis}, {Fukazawa}, {Funk}, {Fusco},
  {Gargano}, {Gasparrini}, {Gehrels}, {Germani}, {Giavitto}, {Giebels},
  {Giglietto}, {Giommi}, {Giordano}, {Giroletti}, {Glanzman}, {Godfrey},
  {Grenier}, {Grondin}, {Grove}, {Guillemot}, {Guiriec}, {Gustafsson},
  {Hadasch}, {Hanabata}, {Harding}, {Hayashida}, {Hays}, {Healey}, {Hill},
  {Horan}, {Hughes}, {Iafrate}, {J{\'o}hannesson}, {Johnson}, {Johnson},
  {Johnson}, {Johnson}, {Kamae}, {Katagiri}, {Kataoka}, {Kawai}, {Kerr},
  {Kn{\"o}dlseder}, {Kocevski}, {Kuss}, {Lande}, {Landriu}, {Latronico}, {Lee},
  {Lemoine-Goumard}, {Lionetto}, {Llena Garde}, {Longo}, {Loparco}, {Lott},
  {Lovellette}, {Lubrano}, {Madejski}, {Makeev}, {Marangelli}, {Marelli},
  {Massaro}, {Mazziotta}, {McConville}, {McEnery}, {Michelson}, {Minuti},
  {Mitthumsiri}, {Mizuno}, {Moiseev}, {Mongelli}, {Monte}, {Monzani},
  {Moretti}, {Morselli}, {Moskalenko}, {Murgia}, {Nakajima}, {Nakamori},
  {Naumann-Godo}, {Nolan}, {Norris}, {Nuss}, {Ohno}, {Ohsugi}, {Omodei},
  {Orlando}, {Ormes}, {Ozaki}, {Paccagnella}, {Paneque}, {Panetta}, {Parent},
  {Pelassa}, {Pepe}, {Pesce-Rollins}, {Pinchera}, {Piron}, {Porter}, {Poupard},
  {Rain{\`o}}, {Rando}, {Ray}, {Razzano}, {Razzaque}, {Rea}, {Reimer},
  {Reimer}, {Reposeur}, {Ripken}, {Ritz}, {Rochester}, {Rodriguez}, {Romani},
  {Roth}, {Sadrozinski}, {Salvetti}, {Sanchez}, {Sander}, {Saz Parkinson},
  {Scargle}, {Schalk}, {Scolieri}, {Sgr{\`o}}, {Shaw}, {Siskind}, {Smith},
  {Smith}, {Spandre}, {Spinelli}, {Starck}, {Stephens}, {Striani}, {Strickman},
  {Strong}, {Suson}, {Tajima}, {Takahashi}, {Takahashi}, {Tanaka}, {Thayer},
  {Thayer}, {Thompson}, {Tibaldo}, {Tibolla}, {Tinebra}, {Torres}, {Tosti},
  {Tramacere}, {Uchiyama}, {Usher}, {Van Etten}, {Vasileiou}, {Vilchez},
  {Vitale}, {Waite}, {Wallace}, {Wang}, {Watters}, {Winer}, {Wood}, {Yang},
  {Ylinen}, {Ziegler}, \& {Fermi LAT Collaboration}}]{1fgl2010}
---. 2010{\natexlab{a}}, \apjs, 188, 405, \dodoi{10.1088/0067-0049/188/2/405}

\bibitem[{{Abdo} {et~al.}(2010{\natexlab{b}}){Abdo}, {Ackermann}, {Ajello},
  {Atwood}, {Axelsson}, {Baldini}, {Ballet}, {Barbiellini}, {Baring},
  {Bastieri}, {Bechtol}, {Bellazzini}, {Berenji}, {Blandford}, {Bloom},
  {Bonamente}, {Borgland}, {Bregeon}, {Brez}, {Brigida}, {Bruel}, {Burnett},
  {Caliandro}, {Cameron}, {Camilo}, {Caraveo}, {Casandjian}, {Cecchi},
  {{\c{C}}elik}, {Chekhtman}, {Cheung}, {Chiang}, {Ciprini}, {Claus},
  {Cognard}, {Cohen-Tanugi}, {Cominsky}, {Conrad}, {Dermer}, {de Angelis}, {de
  Luca}, {de Palma}, {Digel}, {Silva}, {Drell}, {Dubois}, {Dumora}, {Espinoza},
  {Farnier}, {Favuzzi}, {Fegan}, {Ferrara}, {Focke}, {Frailis}, {Freire},
  {Fukazawa}, {Funk}, {Fusco}, {Gargano}, {Gasparrini}, {Gehrels}, {Germani},
  {Giavitto}, {Giebels}, {Giglietto}, {Giordano}, {Glanzman}, {Godfrey},
  {Grenier}, {Grondin}, {Grove}, {Guillemot}, {Guiriec}, {Hanabata}, {Harding},
  {Hayashida}, {Hays}, {Hughes}, {J{\'o}hannesson}, {Johnson}, {Johnson},
  {Johnson}, {Johnson}, {Johnston}, {Kamae}, {Katagiri}, {Kataoka}, {Kawai},
  {Kerr}, {Kn{\"o}dlseder}, {Kocian}, {Kramer}, {Kuehn}, {Kuss}, {Lande},
  {Latronico}, {Lee}, {Lemoine-Goumard}, {Longo}, {Loparco}, {Lott},
  {Lovellette}, {Lubrano}, {Lyne}, {Makeev}, {Marelli}, {Mazziotta}, {McEnery},
  {Meurer}, {Michelson}, {Mitthumsiri}, {Mizuno}, {Moiseev}, {Monte},
  {Monzani}, {Moretti}, {Morselli}, {Moskalenko}, {Murgia}, {Nakamori},
  {Nolan}, {Norris}, {Noutsos}, {Nuss}, {Ohsugi}, {Omodei}, {Orlando}, {Ormes},
  {Ozaki}, {Paneque}, {Panetta}, {Parent}, {Pelassa}, {Pepe}, {Pesce-Rollins},
  {Pierbattista}, {Piron}, {Porter}, {Rain{\`o}}, {Rando}, {Ray}, {Razzano},
  {Reimer}, {Reimer}, {Reposeur}, {Ritz}, {Rochester}, {Rodriguez}, {Romani},
  {Roth}, {Ryde}, {Sadrozinski}, {Sanchez}, {Sander}, {Saz Parkinson},
  {Scargle}, {Sgr{\`o}}, {Siskind}, {Smith}, {Smith}, {Spandre}, {Spinelli},
  {Stappers}, {Strickman}, {Suson}, {Tajima}, {Takahashi}, {Tanaka}, {Thayer},
  {Thayer}, {Theureau}, {Thompson}, {Thorsett}, {Tibaldo}, {Torres}, {Tosti},
  {Tramacere}, {Uchiyama}, {Usher}, {Van Etten}, {Vasileiou}, {Vilchez},
  {Vitale}, {Waite}, {Wallace}, {Wang}, {Watters}, {Weltevrede}, {Winer},
  {Wood}, {Ylinen}, \& {Ziegler}}]{abd+10}
---. 2010{\natexlab{b}}, \apj, 708, 1254, \dodoi{10.1088/0004-637X/708/2/1254}

\bibitem[{{Acero} {et~al.}(2015){Acero}, {Ackermann}, {Ajello}, {Albert},
  {Atwood}, {Axelsson}, {Baldini}, {Ballet}, {Barbiellini}, {Bastieri},
  {Belfiore}, {Bellazzini}, {Bissaldi}, {Blandford}, {Bloom}, {Bogart},
  {Bonino}, {Bottacini}, {Bregeon}, {Britto}, {Bruel}, {Buehler}, {Burnett},
  {Buson}, {Caliandro}, {Cameron}, {Caputo}, {Caragiulo}, {Caraveo},
  {Casandjian}, {Cavazzuti}, {Charles}, {Chaves}, {Chekhtman}, {Cheung},
  {Chiang}, {Chiaro}, {Ciprini}, {Claus}, {Cohen-Tanugi}, {Cominsky}, {Conrad},
  {Cutini}, {D'Ammando}, {de Angelis}, {DeKlotz}, {de Palma}, {Desiante},
  {Digel}, {Di Venere}, {Drell}, {Dubois}, {Dumora}, {Favuzzi}, {Fegan},
  {Ferrara}, {Finke}, {Franckowiak}, {Fukazawa}, {Funk}, {Fusco}, {Gargano},
  {Gasparrini}, {Giebels}, {Giglietto}, {Giommi}, {Giordano}, {Giroletti},
  {Glanzman}, {Godfrey}, {Grenier}, {Grondin}, {Grove}, {Guillemot}, {Guiriec},
  {Hadasch}, {Harding}, {Hays}, {Hewitt}, {Hill}, {Horan}, {Iafrate}, {Jogler},
  {J{\'o}hannesson}, {Johnson}, {Johnson}, {Johnson}, {Johnson}, {Kamae},
  {Kataoka}, {Katsuta}, {Kuss}, {La Mura}, {Landriu}, {Larsson}, {Latronico},
  {Lemoine-Goumard}, {Li}, {Li}, {Longo}, {Loparco}, {Lott}, {Lovellette},
  {Lubrano}, {Madejski}, {Massaro}, {Mayer}, {Mazziotta}, {McEnery},
  {Michelson}, {Mirabal}, {Mizuno}, {Moiseev}, {Mongelli}, {Monzani},
  {Morselli}, {Moskalenko}, {Murgia}, {Nuss}, {Ohno}, {Ohsugi}, {Omodei},
  {Orienti}, {Orlando}, {Ormes}, {Paneque}, {Panetta}, {Perkins},
  {Pesce-Rollins}, {Piron}, {Pivato}, {Porter}, {Racusin}, {Rando}, {Razzano},
  {Razzaque}, {Reimer}, {Reimer}, {Reposeur}, {Rochester}, {Romani},
  {Salvetti}, {S{\'a}nchez-Conde}, {Saz Parkinson}, {Schulz}, {Siskind},
  {Smith}, {Spada}, {Spandre}, {Spinelli}, {Stephens}, {Strong}, {Suson},
  {Takahashi}, {Takahashi}, {Tanaka}, {Thayer}, {Thayer}, {Thompson},
  {Tibaldo}, {Tibolla}, {Torres}, {Torresi}, {Tosti}, {Troja}, {Van Klaveren},
  {Vianello}, {Winer}, {Wood}, {Wood}, {Zimmer}, \& {Fermi-LAT
  Collaboration}}]{3fgl2015}
{Acero}, F., {Ackermann}, M., {Ajello}, M., {et~al.} 2015, \apjs, 218, 23,
  \dodoi{10.1088/0067-0049/218/2/23}

\bibitem[{{Alpar} {et~al.}(1982){Alpar}, {Cheng}, {Ruderman}, \&
  {Shaham}}]{acr+82}
{Alpar}, M.~A., {Cheng}, A.~F., {Ruderman}, M.~A., \& {Shaham}, J. 1982, \nat,
  300, 728, \dodoi{10.1038/300728a0}

\bibitem[{{Archibald} {et~al.}(2009){Archibald}, {Stairs}, {Ransom}, {Kaspi},
  {Kondratiev}, {Lorimer}, {McLaughlin}, {Boyles}, {Hessels}, {Lynch}, {van
  Leeuwen}, {Roberts}, {Jenet}, {Champion}, {Rosen}, {Barlow}, {Dunlap}, \&
  {Remillard}}]{asr+09}
{Archibald}, A.~M., {Stairs}, I.~H., {Ransom}, S.~M., {et~al.} 2009, Science,
  324, 1411, \dodoi{10.1126/science.1172740}

\bibitem[{{Ballet} {et~al.}(2023){Ballet}, {Bruel}, {Burnett}, {Lott}, \& {The
  Fermi-LAT collaboration}}]{4fgl-dr4}
{Ballet}, J., {Bruel}, P., {Burnett}, T.~H., {Lott}, B., \& {The Fermi-LAT
  collaboration}. 2023, arXiv e-prints, arXiv:2307.12546,
  \dodoi{10.48550/arXiv.2307.12546}

\bibitem[{{Benvenuto} {et~al.}(2014){Benvenuto}, {De Vito}, \&
  {Horvath}}]{bdh14}
{Benvenuto}, O.~G., {De Vito}, M.~A., \& {Horvath}, J.~E. 2014, \apjl, 786, L7,
  \dodoi{10.1088/2041-8205/786/1/L7}

\bibitem[{{Bhattacharya} \& {van den Heuvel}(1991)}]{bv91}
{Bhattacharya}, D., \& {van den Heuvel}, E.~P.~J. 1991, \physrep, 203, 1,
  \dodoi{10.1016/0370-1573(91)90064-S}

\bibitem[{{Chen} {et~al.}(2013){Chen}, {Chen}, {Tauris}, \& {Han}}]{cct+13}
{Chen}, H.-L., {Chen}, X., {Tauris}, T.~M., \& {Han}, Z. 2013, \apj, 775, 27,
  \dodoi{10.1088/0004-637X/775/1/27}

\bibitem[{{Clark} {et~al.}(2021){Clark}, {Nieder}, {Voisin}, {Allen},
  {Aulbert}, {Behnke}, {Breton}, {Choquet}, {Corongiu}, {Dhillon},
  {Eggenstein}, {Fehrmann}, {Guillemot}, {Harding}, {Kennedy}, {Machenschalk},
  {Marsh}, {Mata S{\'a}nchez}, {Mignani}, {Stringer}, {Wadiasingh}, \&
  {Wu}}]{cnv+21}
{Clark}, C.~J., {Nieder}, L., {Voisin}, G., {et~al.} 2021, \mnras, 502, 915,
  \dodoi{10.1093/mnras/staa3484}

\bibitem[{{Corbet} {et~al.}(2016){Corbet}, {Chomiuk}, {Coe}, {Coley}, {Dubus},
  {Edwards}, {Martin}, {McBride}, {Stevens}, {Strader}, {Townsend}, \&
  {Udalski}}]{cor+16}
{Corbet}, R.~H.~D., {Chomiuk}, L., {Coe}, M.~J., {et~al.} 2016, \apj, 829, 105,
  \dodoi{10.3847/0004-637X/829/2/105}

\bibitem[{{Corbet} {et~al.}(2019){Corbet}, {Chomiuk}, {Coe}, {Coley}, {Dubus},
  {Edwards}, {Martin}, {McBride}, {Stevens}, {Strader}, \& {Townsend}}]{cor+19}
---. 2019, \apj, 884, 93, \dodoi{10.3847/1538-4357/ab3e32}

\bibitem[{{Corbet} {et~al.}(2022){Corbet}, {Chomiuk}, {Coley}, {Dubus},
  {Edwards}, {Islam}, {McBride}, {Stevens}, {Strader}, {Swihart}, \&
  {Townsend}}]{ccc+22}
{Corbet}, R.~H.~D., {Chomiuk}, L., {Coley}, J.~B., {et~al.} 2022, \apj, 935, 2,
  \dodoi{10.3847/1538-4357/ac6fe2}

\bibitem[{{Corongiu} {et~al.}(2021){Corongiu}, {Mignani}, {Seyffert}, {Clark},
  {Venter}, {Nieder}, {Possenti}, {Burgay}, {Belfiore}, {De Luca}, {Ridolfi},
  \& {Wadiasingh}}]{cms+21}
{Corongiu}, A., {Mignani}, R.~P., {Seyffert}, A.~S., {et~al.} 2021, \mnras,
  502, 935, \dodoi{10.1093/mnras/staa3463}

\bibitem[{{Fermi LAT Collaboration} {et~al.}(2012){Fermi LAT Collaboration},
  {Ackermann}, {Ajello}, {Ballet}, {Barbiellini}, {Bastieri}, {Belfiore},
  {Bellazzini}, {Berenji}, {Blandford}, {Bloom}, {Bonamente}, {Borgland},
  {Bregeon}, {Brigida}, {Bruel}, {Buehler}, {Buson}, {Caliandro}, {Cameron},
  {Caraveo}, {Cavazzuti}, {Cecchi}, {{\c{C}}elik}, {Charles}, {Chaty},
  {Chekhtman}, {Cheung}, {Chiang}, {Ciprini}, {}, {Claus}, {Cohen-Tanugi},
  {Corbel}, {Corbet}, {Cutini}, {de Luca}, {den Hartog}, {de Palma}, {Dermer},
  {Digel}, {do Couto e Silva}, {Donato}, {Drell}, {Drlica-Wagner}, {Dubois},
  {Dubus}, {Favuzzi}, {Fegan}, {Ferrara}, {Focke}, {Fortin}, {Fukazawa},
  {Funk}, {Fusco}, {Gargano}, {Gasparrini}, {Gehrels}, {Germani}, {Giglietto},
  {Giordano}, {Giroletti}, {Glanzman}, {Godfrey}, {Grenier}, {Grove},
  {Guiriec}, {Hadasch}, {Hanabata}, {Harding}, {Hayashida}, {Hays}, {Hill},
  {Hughes}, {J{\'o}hannesson}, {Johnson}, {Johnson}, {Kamae}, {Katagiri},
  {Kataoka}, {Kerr}, {Kn{\"o}dlseder}, {Kuss}, {Lande}, {Longo}, {Loparco},
  {Lovellette}, {Lubrano}, {Mazziotta}, {McEnery}, {Michelson}, {Mitthumsiri},
  {Mizuno}, {Monte}, {Monzani}, {Morselli}, {Moskalenko}, {Murgia}, {Nakamori},
  {Naumann-Godo}, {Norris}, {Nuss}, {Ohno}, {Ohsugi}, {Okumura}, {Omodei},
  {Orlando}, {Ozaki}, {Paneque}, {Parent}, {Pesce-Rollins}, {Pierbattista},
  {Piron}, {Pivato}, {Porter}, {Rain{\`o}}, {Rando}, {Razzano}, {Reimer},
  {Reimer}, {Ritz}, {Romani}, {Roth}, {Saz Parkinson}, {Sgr{\`o}}, {Siskind},
  {Spandre}, {Spinelli}, {Suson}, {Takahashi}, {Tanaka}, {Thayer}, {Thayer},
  {Thompson}, {Tibaldo}, {Tinivella}, {Torres}, {Tosti}, {Troja}, {Uchiyama},
  {Usher}, {Vandenbroucke}, {Vianello}, {Vitale}, {Waite}, {Winer}, {Wood},
  {Wood}, {Yang}, {Zimmer}, {Coe}, {Di Mille}, {Edwards}, {Filipovi{\'c}},
  {Payne}, {Stevens}, \& {Torres}}]{lat+12}
{Fermi LAT Collaboration}, {Ackermann}, M., {Ajello}, M., {et~al.} 2012,
  Science, 335, 189, \dodoi{10.1126/science.1213974}

\bibitem[{{Fruchter} {et~al.}(1988){Fruchter}, {Stinebring}, \&
  {Taylor}}]{fst88}
{Fruchter}, A.~S., {Stinebring}, D.~R., \& {Taylor}, J.~H. 1988, \nat, 333,
  237, \dodoi{10.1038/333237a0}

\bibitem[{{Kerr}(2011)}]{k11}
{Kerr}, M. 2011, \apj, 732, 38, \dodoi{10.1088/0004-637X/732/1/38}

\bibitem[{{Lomb}(1976)}]{l76}
{Lomb}, N.~R. 1976, \apss, 39, 447, \dodoi{10.1007/BF00648343}

\bibitem[{{Ng} {et~al.}(2018){Ng}, {Takata}, {Strader}, {Li}, \&
  {Cheng}}]{nts+18}
{Ng}, C.~W., {Takata}, J., {Strader}, J., {Li}, K.~L., \& {Cheng}, K.~S. 2018,
  \apj, 867, 90, \dodoi{10.3847/1538-4357/aae308}

\bibitem[{{Nolan} {et~al.}(2012){Nolan}, {Abdo}, {Ackermann}, {Ajello},
  {Allafort}, {Antolini}, {Atwood}, {Axelsson}, {Baldini}, {Ballet},
  {Barbiellini}, {Bastieri}, {Bechtol}, {Belfiore}, {Bellazzini}, {Berenji},
  {Bignami}, {Blandford}, {Bloom}, {Bonamente}, {Bonnell}, {Borgland},
  {Bottacini}, {Bouvier}, {Brandt}, {Bregeon}, {Brigida}, {Bruel}, {Buehler},
  {Burnett}, {Buson}, {Caliandro}, {Cameron}, {Campana}, {Ca{\~n}adas},
  {Cannon}, {Caraveo}, {Casandjian}, {Cavazzuti}, {Ceccanti}, {Cecchi},
  {{\c{C}}elik}, {Charles}, {Chekhtman}, {Cheung}, {Chiang}, {Chipaux},
  {Ciprini}, {Claus}, {Cohen-Tanugi}, {Cominsky}, {Conrad}, {Corbet}, {Cutini},
  {D'Ammando}, {Davis}, {de Angelis}, {DeCesar}, {DeKlotz}, {De Luca}, {den
  Hartog}, {de Palma}, {Dermer}, {Digel}, {Silva}, {Drell}, {Drlica-Wagner},
  {Dubois}, {Dumora}, {Enoto}, {Escande}, {Fabiani}, {Falletti}, {Favuzzi},
  {Fegan}, {Ferrara}, {Focke}, {Fortin}, {Frailis}, {Fukazawa}, {Funk},
  {Fusco}, {Gargano}, {Gasparrini}, {Gehrels}, {Germani}, {Giebels},
  {Giglietto}, {Giommi}, {Giordano}, {Giroletti}, {Glanzman}, {Godfrey},
  {Grenier}, {Grondin}, {Grove}, {Guillemot}, {Guiriec}, {Gustafsson},
  {Hadasch}, {Hanabata}, {Harding}, {Hayashida}, {Hays}, {Hill}, {Horan},
  {Hou}, {Hughes}, {Iafrate}, {Itoh}, {J{\'o}hannesson}, {Johnson}, {Johnson},
  {Johnson}, {Johnson}, {Kamae}, {Katagiri}, {Kataoka}, {Katsuta}, {Kawai},
  {Kerr}, {Kn{\"o}dlseder}, {Kocevski}, {Kuss}, {Lande}, {Landriu},
  {Latronico}, {Lemoine-Goumard}, {Lionetto}, {Llena Garde}, {Longo},
  {Loparco}, {Lott}, {Lovellette}, {Lubrano}, {Madejski}, {Marelli}, {Massaro},
  {Mazziotta}, {McConville}, {McEnery}, {Mehault}, {Michelson}, {Minuti},
  {Mitthumsiri}, {Mizuno}, {Moiseev}, {Mongelli}, {Monte}, {Monzani},
  {Morselli}, {Moskalenko}, {Murgia}, {Nakamori}, {Naumann-Godo}, {Norris},
  {Nuss}, {Nymark}, {Ohno}, {Ohsugi}, {Okumura}, {Omodei}, {Orlando}, {Ormes},
  {Ozaki}, {Paneque}, {Panetta}, {Parent}, {Perkins}, {Pesce-Rollins},
  {Pierbattista}, {Pinchera}, {Piron}, {Pivato}, {Porter}, {Racusin},
  {Rain{\`o}}, {Rando}, {Razzano}, {Razzaque}, {Reimer}, {Reimer}, {Reposeur},
  {Ritz}, {Rochester}, {Romani}, {Roth}, {Rousseau}, {Ryde}, {Sadrozinski},
  {Salvetti}, {Sanchez}, {Saz Parkinson}, {Sbarra}, {Scargle}, {Schalk},
  {Sgr{\`o}}, {Shaw}, {Shrader}, {Siskind}, {Smith}, {Spandre}, {Spinelli},
  {Stephens}, {Strickman}, {Suson}, {Tajima}, {Takahashi}, {Takahashi},
  {Tanaka}, {Thayer}, {Thayer}, {Thompson}, {Tibaldo}, {Tibolla}, {Tinebra},
  {Tinivella}, {Torres}, {Tosti}, {Troja}, {Uchiyama}, {Vandenbroucke}, {Van
  Etten}, {Van Klaveren}, {Vasileiou}, {Vianello}, {Vitale}, {Waite},
  {Wallace}, {Wang}, {Werner}, {Winer}, {Wood}, {Wood}, {Wood}, {Yang}, \&
  {Zimmer}}]{2fgl2012}
{Nolan}, P.~L., {Abdo}, A.~A., {Ackermann}, M., {et~al.} 2012, \apjs, 199, 31,
  \dodoi{10.1088/0067-0049/199/2/31}

\bibitem[{{Ravi} {et~al.}(2010){Ravi}, {Manchester}, \& {Hobbs}}]{rmh10}
{Ravi}, V., {Manchester}, R.~N., \& {Hobbs}, G. 2010, \apjl, 716, L85,
  \dodoi{10.1088/2041-8205/716/1/L85}

\bibitem[{{Roberts}(2013)}]{rob13}
{Roberts}, M. S.~E. 2013, in IAU Symposium, Vol. 291, Neutron Stars and
  Pulsars: Challenges and Opportunities after 80 years, ed. J.~{van Leeuwen},
  127--132, \dodoi{10.1017/S174392131202337X}

\bibitem[{{Salvetti} {et~al.}(2015){Salvetti}, {Mignani}, {De Luca}, {Delvaux},
  {Pallanca}, {Belfiore}, {Marelli}, {Breeveld}, {Greiner}, {Becker}, \&
  {Pizzocaro}}]{smd+15}
{Salvetti}, D., {Mignani}, R.~P., {De Luca}, A., {et~al.} 2015, \apj, 814, 88,
  \dodoi{10.1088/0004-637X/814/2/88}

\bibitem[{{Scargle}(1982)}]{s82}
{Scargle}, J.~D. 1982, \apj, 263, 835, \dodoi{10.1086/160554}

\bibitem[{{Sim} {et~al.}(2024){Sim}, {An}, \& {Wadiasingh}}]{saw24}
{Sim}, M., {An}, H., \& {Wadiasingh}, Z. 2024, \apj, 964, 109,
  \dodoi{10.3847/1538-4357/ad25fb}

\bibitem[{{Wu} {et~al.}(2012){Wu}, {Takata}, {Cheng}, {Huang}, {Hui}, {Kong},
  {Tam}, \& {Wu}}]{Wu2012}
{Wu}, E.~M.~H., {Takata}, J., {Cheng}, K.~S., {et~al.} 2012, \apj, 761, 181,
  \dodoi{10.1088/0004-637X/761/2/181}

\bibitem[{{Zechmeister} \& {K{\"u}rster}(2009)}]{zk09}
{Zechmeister}, M., \& {K{\"u}rster}, M. 2009, \aap, 496, 577,
  \dodoi{10.1051/0004-6361:200811296}

\end{thebibliography}
\end{document}